\documentclass{kluwer}
\usepackage{klucite}
\usepackage[dvips]{graphicx}
\def\ie {$i.e.,\;$}

\def\ie {$i.e.,\;$}

\def\viz {$viz.,\;$}

\def\La1215 {Ly$\alpha\lambda1215$}

\begin{document}
\begin{article}
\begin{opening}
\title{Magnetic Field Geometry in ``Red" and ``Blue" BL~Lacs}
\author{P. \surname{Kharb}\email{rhea@iiap.ernet.in}}
\institute{Indian Institute of Astrophysics, Bangalore 560 034, India}
\author{D.C. \surname{Gabuzda}}
\institute{Physics Department, University College Cork, Cork, Ireland}
\author{P. \surname{Shastri}}
\institute{Indian Institute of Astrophysics, Bangalore 560 034, India}
\runningtitle{Magnetic Field Geometry in ``Red" and ``Blue" BL~Lacs}
\runningauthor{Kharb e.a.}

\begin{abstract}
We compare the systematics of the magnetic field geometry
in the ``red" low-energy peaked BL Lacs (LBLs) and 
``blue" high-energy peaked BL Lacs
(HBLs) using VLBI polarimetric images. The LBLs are
primarily ``radio--selected" BL Lacs and the HBLs are
primarily ``X-ray selected". In contrast to the LBLs, which
show predominantly transverse jet magnetic fields, the HBLs show
predominantly longitudinal fields. Thus, while the SED peaks of
core-dominated quasars, LBLs and HBLs form a sequence of increasing
frequency, the magnetic field geometry does not follow an analogous
sequence.  We briefly investigate possible connections between the
observed parsec-scale magnetic field structures and circular
polarization measurements in the literature on various spatial
scales.

\end{abstract}
\end{opening}

\section{Introduction}
The central regions of Active Galactic Nuclei (AGN) are far more powerful
emitters than all the stars of the galaxy combined.
AGN classified as BL~Lac objects are characterized by a
predominantly non-thermal, highly polarized continuum that is
variable in both total intensity and polarization at all observed wavelengths,
with weak or no optical line emission. All but the last characteristic are
shared by Optically Violently Variable (OVV) quasars. These extreme
phenomena are understood as a consequence of relativistic beaming in their
nuclei.

The near-consensus view of AGN is that they are powered by mass flows
around a super-massive black hole.
The details of the physical processes are still not well-understood, however,
and we do not yet have a comprehensive theory of AGN that can predict the
whole range and variety of observed AGN properties from a minimal set of
well-defined parameters. Therefore, taxonomy as a substitute for quantification
of parameters still plays a major role in the investigation of their physics.
For example, radio galaxies are classified into types FR~I and II since
there are no known physical parameter(s) that can explain the whole
range in properties of radio galaxies as a class.

\begin{figure*}[h]
\includegraphics[width=11.5cm,height=3.5cm]{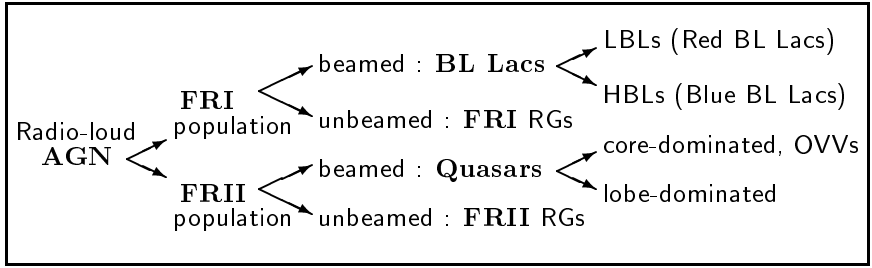}
\caption{The adopted AGN taxonomy.}
\end{figure*}

In the ``radio--loud" AGN ({\it i.e.,} those AGNs with radio power exceeding
the optical power by an order of magnitude or more)
there is compelling evidence that bulk relativistic motion 
\cite{BlandfordKonigl79} is ubiquitous, which leads to relativistic
aberration effects, so that the angle to the line of sight plays a dominant
role in the appearance of these objects \cite{UrryPadovani95}. 
For every Doppler-boosted object, there are intrinsically similar objects 
whose jets are not directed toward us -- the ``parent" population. By
comparing properties that are known to be independent of orientation for 
objects that are known to be boosted and not boosted, we can attempt to 
identify a given class of Doppler-beamed objects with their plane-of-sky 
counterparts or ``parent objects".
For example, BL~Lac objects are believed to be beamed FR~I radio galaxies
\cite{OrrBrowne82,Wardle84}, \ie radio galaxies with relatively
low radio luminosities and jets that spread out into diffuse plumes
\cite{FanaroffRiley74}.
Similarly, quasars have been identified as beamed FR~II radio
galaxies (the OVVs being the most highly beamed ones). This process of 
identifying beamed and parent objects is known as Unification of AGNs.
Whether or not there is a dichotomy in the intrinsic properties of
FR~I and FR~II radio galaxies remains a matter of debate, as is the
physical origin of the observed differences.

\section{``Red" and ``Blue" BL~Lacs}

Among the BL~Lacs, two subclasses have emerged.
The first BL~Lacs were discovered from radio surveys, but X-ray surveys
later yielded many more BL~Lacs which had somewhat differing properties.
The radio-selected BL~Lacs ({\bf RBLs}) were typically more core-dominated on
arcsec-scales, showed higher average optical polarization and greater
variability and had more powerful radio lobes than the X-ray selected
BL~Lacs ({\bf XBLs}) \cite{LaurentMuehleisen93}.
In other words, it appeared that the RBLs were more ``extreme'' than XBLs.

In the light of the fact that orientation is known to
play a major role in the observed properties of AGN, it was suggested that
RBLs were oriented at closer angles to the line of sight than XBLs --
the ``Different Angle Scenario" \cite{Stocke85}. This was explained in
the framework of an accelerating jet model \cite{Ghisellini89},
wherein the X-rays were
radiated in the slower part of the jet and were therefore less
beamed than the radio photons coming from the faster portion of the
jet. The different-angle scenario predicted  that XBLs should be more
numerous than RBLs in a statistically complete sample. The prediction was
consistent with the observations of the time.

The fly in the ointment was the systematic trend discovered in the spectral
energy distributions ({\bf SEDs}) (Figure 2, Urry and Padovani (1995)).
The RBL SEDs typically peak in the near infrared
(low--energy peaked BL Lacs ({\bf LBLs}) or ``red" BL Lacs),
whereas the XBL SEDs typically peak in the soft X-ray regime
(high--energy peaked BL Lacs ({\bf HBLs}) or ``blue" BL Lacs). 
The different-angle scenario could not explain this difference in the SED 
peaks.
Most, though not all, RBLs are LBLs and most, though not all, XBLs are HBLs.
We will adopt this latter terminology, since it is more physical.

\begin{figure*}[h]
\centerline{\includegraphics[width=9.5cm]{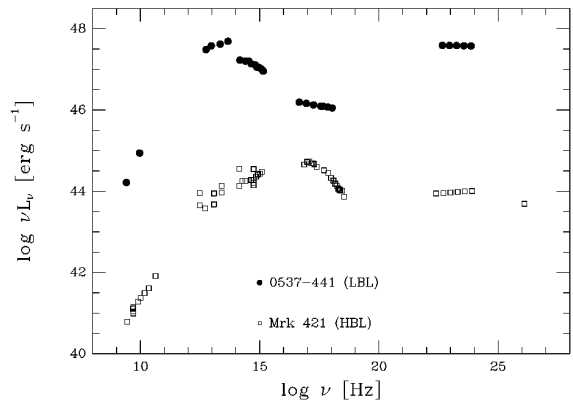}}
\caption{Spectral Energy Distributions show that the synchrotron peaks for
the LBLs lie in the NIR/optical regime and for HBLs in the EUV/soft-X-ray
regime.}
\end{figure*}

The discovery of the systematic differences in the SED peaks of the ``blue"
HBLs and ``red" LBLs gave rise to the alternative ``Different Energy Cutoff 
Scenario," wherein the X-ray emitting regions of HBLs have higher electron 
Lorentz factors or magnetic fields than LBLs \cite{Sambruna96}. This scenario
would then predict that LBLs should be more numerous than HBLs.

\section{Motivation} 

The peaks of the SEDs of core-dominated quasars (CDQs), LBLs and HBLs
form a sequence of increasing frequency. The CDQs are thought to be beamed
FR~II radio galaxies, while the BL Lac objects are thought to be beamed
FR~I radio galaxies. As if reflecting this dichotomy, there is a systematic difference between
the magnetic-field geometries observed for the VLBI jets of BL~Lacs (mostly
LBLs) and CDQs, \viz the ordered component of the jet magnetic field tends
to be parallel (perpendicular) to the local jet direction in CDQs (LBLs)
\cite{Cawthorne93, Gabuzda00}.

We wished to explore the parsec-scale magnetic field geometries in the HBLs,
to see if this geometry reflected the sequence in the SED peaks. We
accordingly undertook pc-scale polarimetric imaging of X-ray selected BL Lacs
from the HEAO-1 survey.

\section{Observations}

Polarization VLBI observations were made at 5~GHz in February 1993 (using
a global array of three European and six US antennas), July 1995 and June
1998 (both using the NRAO 
\footnote{The National Radio Astronomy Observatory is a facility of the 
National Science Foundation operated under cooperative agreement by 
Associated Universities, Inc.} Very Large Baseline Array). 
The sample consisted of 21 northern 
hemisphere BL Lacs detected in hard X-rays by the {\it HEAO}--1 survey 
\cite{Wood84} and six BL Lacs from the {\it ROSAT}-Green Bank (RGB)
sample \cite{LaurentMuehleisen97}. Calibration, fringe fitting and 
imaging were done using the Astronomical Image Processing System (AIPS) 
with the polarization calibration following standard methods \cite{Cotton89}. 
Results for 4 XBLs from this sample have been previously published 
\cite{Kollgaard96}. 

\section{Results}

The VLBI polarization observations of LBLs (primarily from the 1-Jy sample
\cite{KuehrSchmidt90}, {\it e.g.}, Gabuzda et al. (2000) and
references therein) and HBLs (our work) reveal that both LBLs and HBLs have
compact ``core-jet" morphologies. We present the radio maps of some of the
objects from the HEAO-1 sample in Figures~3--4.

Several of the HBLs that we have observed show evidence for a
``spine+sheath" magnetic-field structure, with the inner region if the jet
having transverse {\bf B} field and the edges having longitudinal magnetic
field. This type of {\bf B}-field structure could be a result of interaction
of the jet with the surrounding medium. Particularly good
examples are the HBLs 1230+253 (Figure 4c) and 1727+502 (Figure 4d). This
may point in the direction of a helical magnetic field threading the jets of
at least some of these objects, which could have implications for mechanisms
for the production of circular polarization in them.

\begin{figure}[h]
(a)
\centerline{\includegraphics[width=8cm]{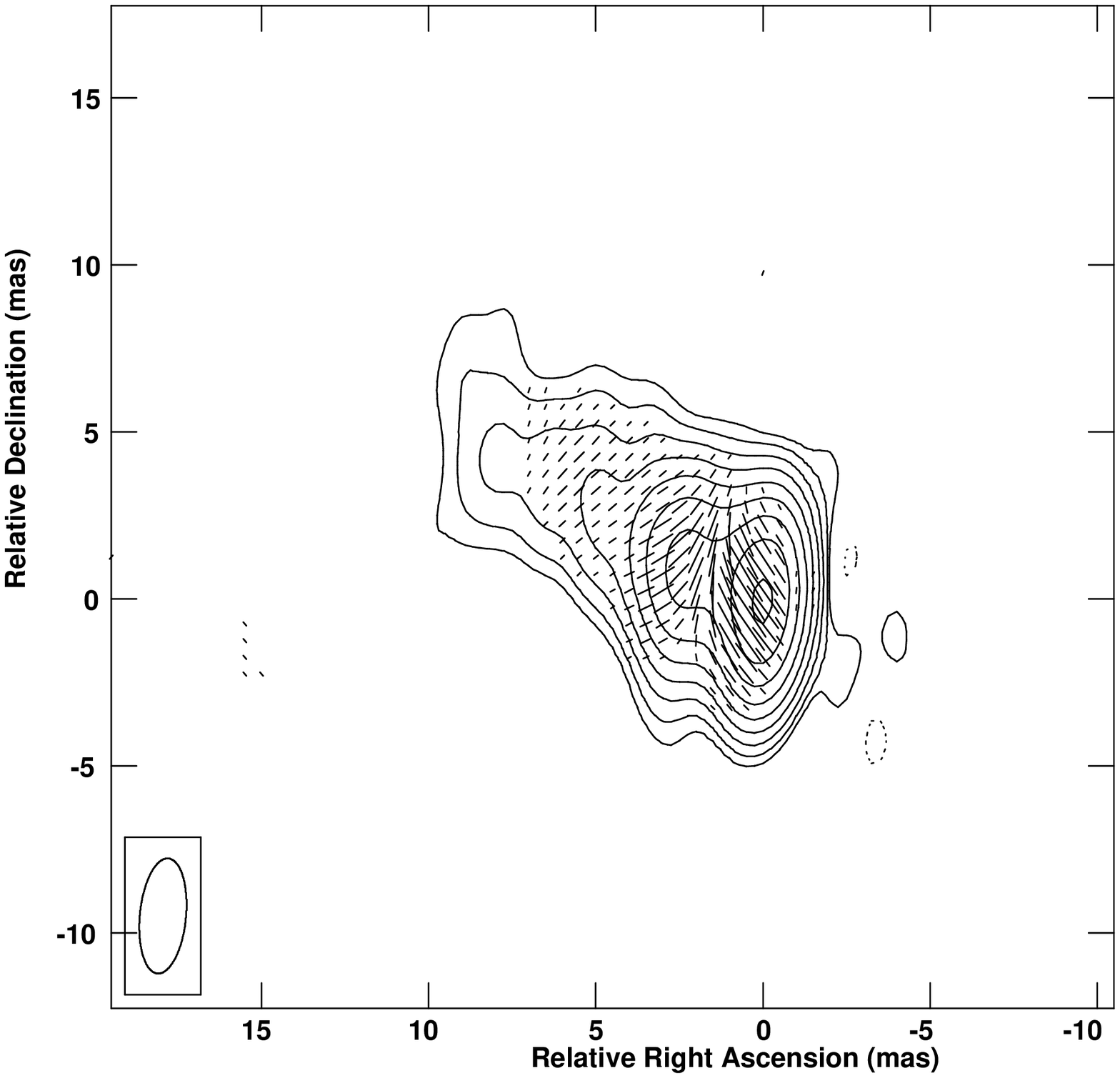}}
(b)
\centerline{\includegraphics[width=8cm]{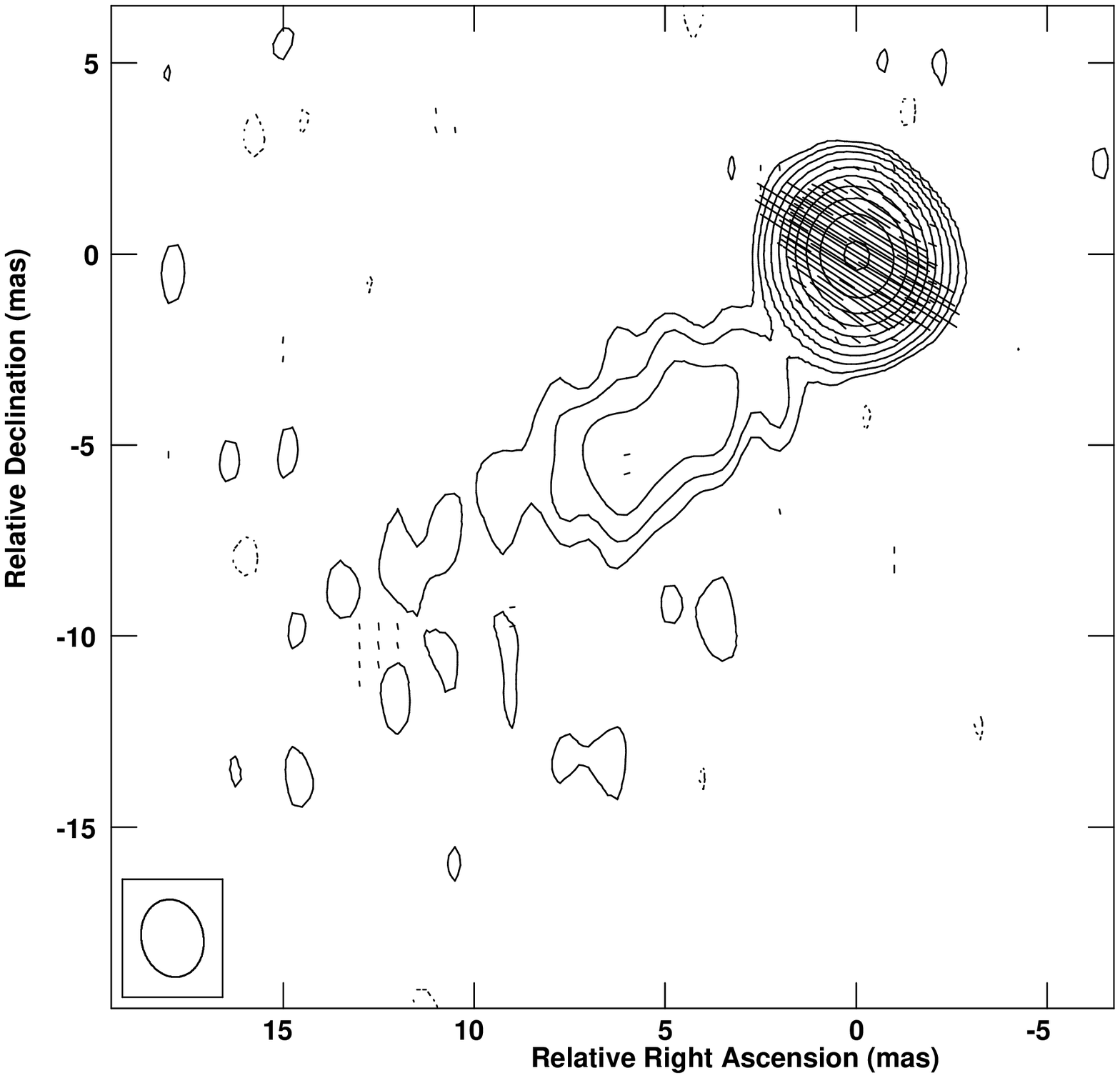}}
\caption{Total intensity VLBI maps of LBLs (a) 0829+046 and (b)
0929+502 at 5 GHz with polarization electric vectors superimposed. 
Contours are (a) -0.35, 0.35, 0.70, 1.40, 2.80, 5.60, 11.20, 22.50, 45 
and 90 per cent of the peak brightness of 441 mJy beam$^{-1}$,
$\chi$ vectors: 1 mas = 8 mJy beam$^{-1}$ and (b)
-0.17, 0.17, 0.35, 0.70, 1.40, 2.80, 5.60, 11.20, 22.50, 45 and 90
per cent of the peak brightness of 433 mJy beam$^{-1}$, $\chi$ vectors: 
1 mas = 10 mJy beam$^{-1}$.}
\end{figure}

\begin{figure}[h]
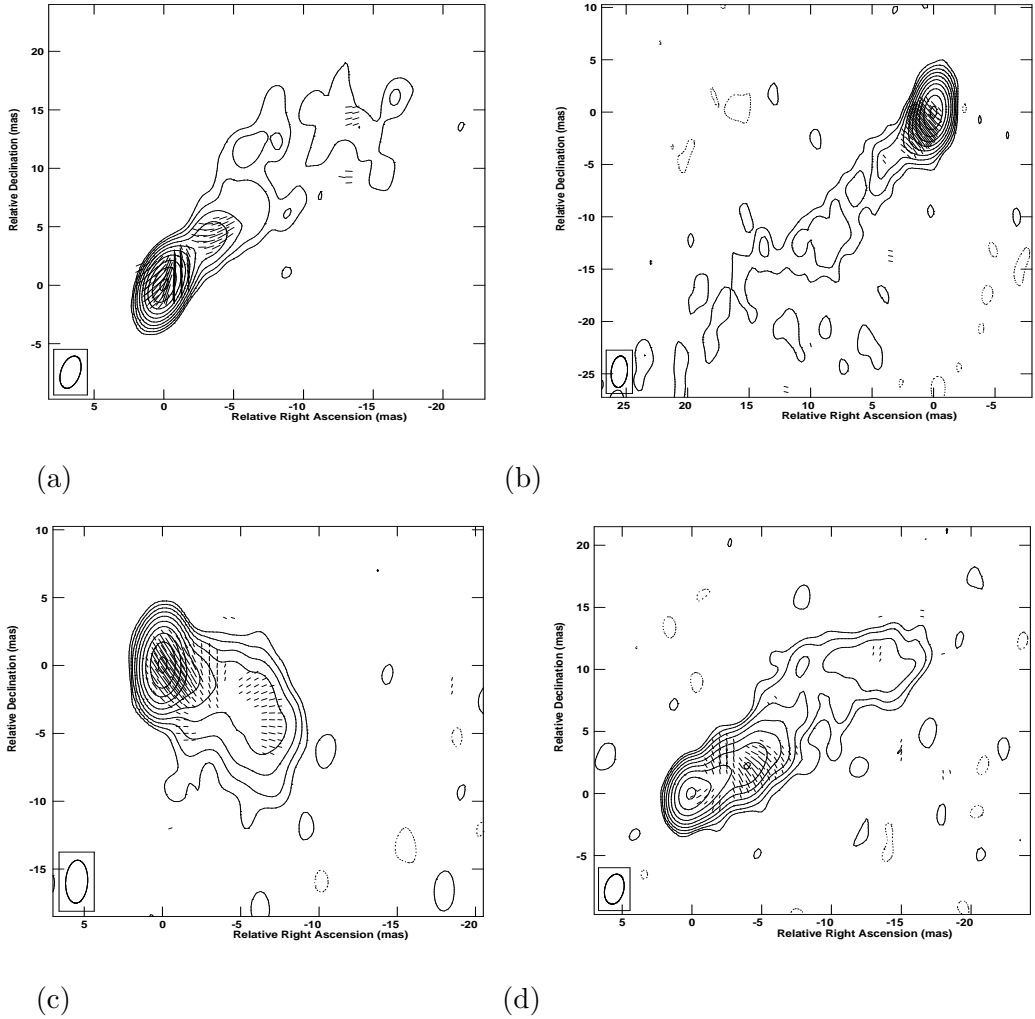

(a)
\centerline{\includegraphics[width=6.5cm,height=6.8cm]{KharbShastri_5.ps}
(b)
\includegraphics[width=6.5cm,height=6.8cm]{KharbShastri_6.ps}}
(c)
\centerline{\includegraphics[width=6.5cm,height=6.8cm]{KharbShastri_7.ps}
(d)
\includegraphics[width=6.5cm,height=6.8cm]{KharbShastri_8.ps}}
\caption{Total intensity VLBI maps of HBLs (a) 1101+384 (b) 1215+303
(c) 1230+253 and (d) 1727+502 at 5 GHz with $\chi$ vectors superimposed.
Contours are (a) -0.17, 0.17, 0.35, 0.70, 1.40, 2.80, 5.70, 11.50, 22.50, 45
and 90 per cent of the peak brightness of 356 mJy beam$^{-1}$, 
$\chi$ vectors: 1 mas = 1.8 mJy beam$^{-1}$ (b) -0.17,
0.17, 0.35, 0.70, 1.40, 2.80, 5.70, 11.50, 22.50, 45 and 90 percent
of the peak brightness of 231 mJy beam$^{-1}$, $\chi$ vectors: 1 mas =
2 mJy beam$^{-1}$, (c) -0.17, 0.17, 0.35, 0.70,
1.40, 2.80, 5.60, 11.20, 22.50, 45 and 90 percent of the peak brightness of
183 mJy beam$^{-1}$, $\chi$ vectors: 1 mas = 2.5 mJy beam$^{-1}$ and 
(d) -0.35, 0.35, 0.70, 1.40, 2.80, 5.60, 11.20, 
22.50, 45 and 90 per cent of the peak brightness of 64 mJy beam$^{-1}$,
$\chi$ vectors: 1 mas = 1.8 mJy beam$^{-1}$.} 
\end{figure}

The orientation of the parsec-scale core polarizations
of HBLs and LBLs relative to the inner VLBI jet direction do not show
any obvious systematic differences. In both cases, there is a clear tendency
for the core polarization electric vectors to lie either parallel or
perpendicular to the jet direction \cite{Gabuzda00}. The physical origin
of this bimodal behaviour is not entirely clear, though it may be due at
least in part to optical depth effects (Gabuzda, these proceedings).

In contrast, the {\em jet} polarizations of HBLs and LBLs do display
systematic differences, as can be seen in Table~I. Among both types of BL Lac
there are sources in which the observed jet polarization angles lie
parallel or perpendicular to the local jet direction. However, the
polarization angles are perpendicular to the jet in the majority of HBLs,
while they are aligned with the jet in the majority of LBLs. Assuming that
the emission region is optically thin, which is certainly the case for the jet,
we infer that the magnetic ({\bf B}) field is perpendicular to the observed
polarization angle.
Thus, the LBLs show predominantly {\em transverse} jet magnetic fields,
while the HBLs show predominantly {\em longitudinal} jet fields.

\begin{table}[h]
\begin{tabular*}{\maxfloatwidth}{lcc}
\hline
~~~~~{\bf B field structure in the VLBI jet}  & {\bf LBL} &{\bf HBL}
 \\ \hline
{\small {\bf B} is transverse
to local jet} & 65 \% & 35 \%\\
{\small {\bf B} is parallel
to local jet} & 35 \% & 55 \% \\
{\small No obvious relation}
 & -- & 10 \% \\ \hline
\end{tabular*}
\caption{Systematics of B field geometry in the two subclasses
of BL Lacs.}
\end{table}

Since the jets of CDQs exhibit predominatly {\em longitudinal} magnetic
fields, we find that the systematics of the jet {\bf B} field
geometry in CDQs, LBLs and HBLs, which goes from longitudinal to transverse
to longitudinal, does not reflect the sequence in the SED peaks.

\section{Circular polarization}

It has been argued that the sequence in the SED peaks of CDQs, LBLs
and HBLs might reflect a sequence in magnetic field strength/Lorentz factors
\cite{Sambruna94}.
Since the detectability of circular polarization may also be linked to the
magnetic field strength and/or geometry and the low-end of the Lorentz factor
distrubution (if the circular polarization is produced by Faraday conversion
of linear polarization), we looked for evidence for a connection between the
parsec-scale {\bf B} field geometry and previous detections of circular
polarization.

To this end we compiled all previous circular polarization measurements we
could find in the literature (see Table II). We found such measurements for
a total of 73 quasars and 21 BL~Lacs, in addition to 52 galaxies, made on
various spatial scales. Of the 21 BL~Lacs, 19 are LBLs and only 2 are HBLs.
We divided the measurements into three groups on the basis of spatial scale:
parsec-scale, kiloparsec-scale ATCA, and kiloparsec-scale single-dish
(Table II).

\begin{table}[h]
\begin{tabular*}{\maxfloatwidth}{lccc}
\hline
& & Detection rate \% (no. observed)\\
& Single-dish (4, 5) & ATCA (3) & VLBI (1, 2) \\ 
& (kpc-scale) & (kpc-scale) & (pc-scale) \\\hline
{\bf Quasars} & 46 \% (52) & 67 \% (15) & 23 \% (35) \\
{\bf BL Lacs } & 27 \% (11) & 60 \% (5)  &  0  \% (12) \\ \hline
\end{tabular*}
\caption{Circular polarization detection rates in Quasars and BL Lacs.}
{References : 1: \cite{HomanWardle99}, 2: \cite{HomanAttridgeWardle01},
3: \cite{Rayner00}, 4: \cite{WeilerDePater83}, 5: \cite{Komesaroff84}}
\end{table}

The statistics are insufficient to discern differences between the
circular-polarization detection rates for LBLs and HBLs, since only 2 HBLs
were observed in the circular polarization experiments.
There is some evidence from the VLBI measurements (Homan et al., 2001) and
single-dish measurements (Weiler and de Pater, 1983; Komesaroff et al., 1984)
that the circular-polarization detection rate may be higher for quasars
than for BL~Lacs. 
Could this be a consequence of the fact that most of the quasars
observed have brighter compact nuclei than the BL~Lacs~? 

Figure 5 shows the distribution of the total flux density on the {\it relevant}
spatial scales for all the circular polarization observations. 
The observations are again divided into groups as in Table II (see above).
In the case of the single-dish measurements of circular polarization (left-most
panels in each row), the distributions are of the total flux density of 
the {\it compact} component, {\it i.e.,} from a VLBI measurement, 
as given in Weiler and de Pater, 1983. In the case of the ATCA and VLBI 
measurements (middle and right-most panels), the distributions are of the 
total intensity measured by ATCA and VLBI, respectively.  It is especially
clear in the VLBI circular polarization measurements (last column of 
histograms) that nearly all
of the relatively few BL Lac objects for which circular polarization
measurements are available fall at the low end of the flux density range
for the observed quasars. However, the situation is not clear, since (i) there
is no obvious difference in the detection rates for CDQs and BL Lacs in the
ATCA measurements (Rayner et al., 2000), and (ii) the quasars in which
circular polarization is detected in the ATCA and single-dish observations
have flux densities  that fall in the same range as those of BL Lac objects.
Further, in the single-dish observations of radio galaxies, circular 
polarization is detected in compact components with much lower flux 
densities than the weakest components in BL Lacs in which circular 
polarization is detected.

\begin{figure*}[h]
\centerline{\includegraphics{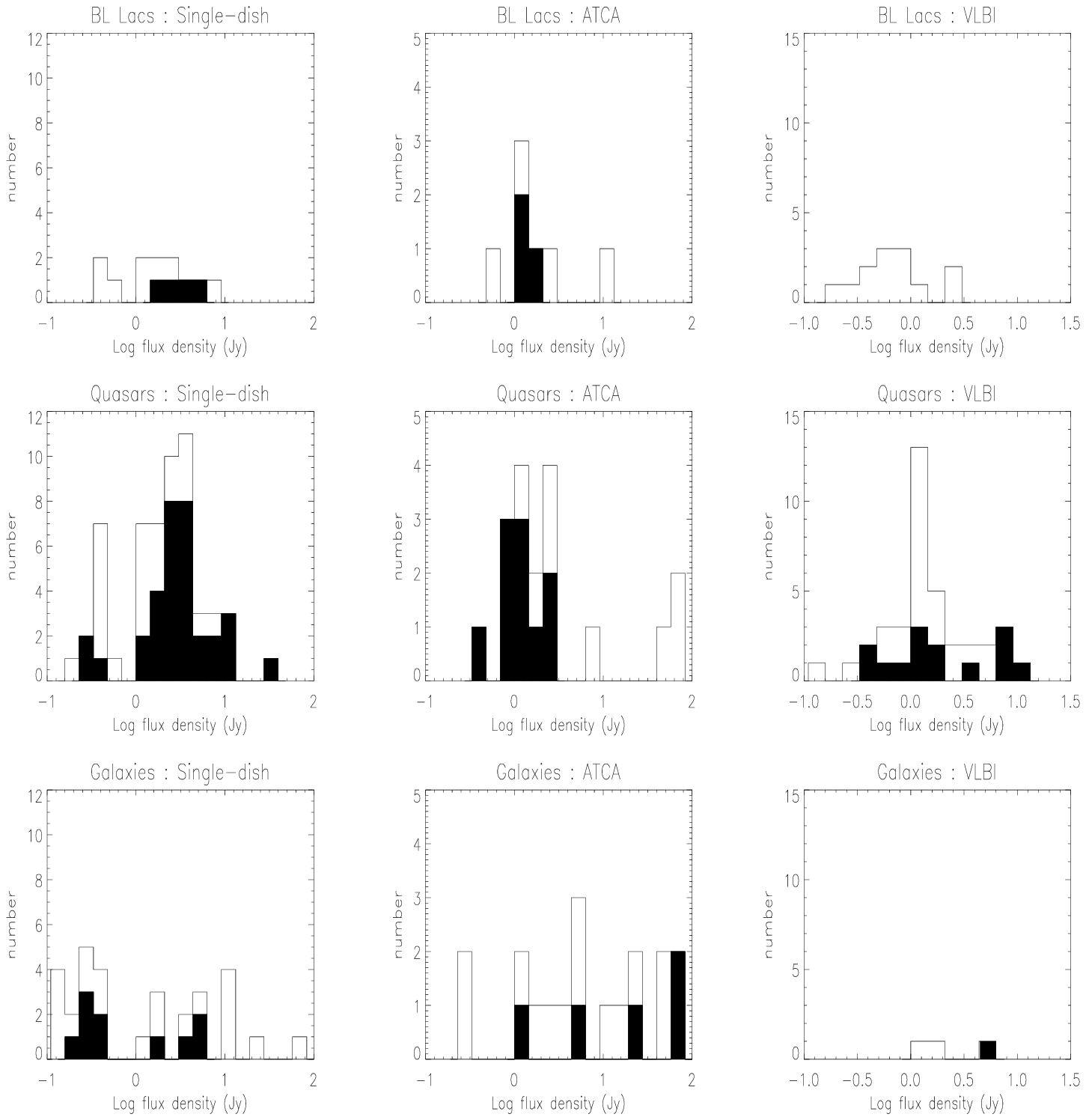}}
\caption{The histograms showing the number of observed
and detected (shaded black) BL Lac objects, quasars and radio galaxies with 
respect to the total flux density, at three different spatial scales for 
circular polarization (see text for details), with resolution increasing from 
left to right panels in each row.}
\end{figure*}

If systematic differences in the circular-polarization detection
rates (or other properties of the circular polarization) observed in CDQs,
LBLs and HBLs could reliably be shown to be present or absent, this could
lead to interesting clues to the origin of the circular polarization.
Circular polarization observations of well-selected samples, for {\it eg.,}
comparable HBLs and LBLs, are needed
to search for relationships between magnetic-field geometry and other
properties of the compact radio emission and the circular-polarization
detection rate. Well-selected samples might necessitate observations
of relatively faint sources, but this is not unrealistic as Figure 5 shows.
If established, such relationships could potentially
provide useful information about the Lorentz factors and magnetic-field
strengths in the jets of different subclasses of BL~Lacs and quasars.

A circular polarization analysis of all sources in the 1-Jy sample of northern 
BL Lac objects (mostly LBLs) defined by Kuehr and Schmidt (1990) is underway
(Gabuzda and Vitrischak, in preparation).

\section{Conclusions}

\begin{enumerate}
\item The ordered component of the {\bf B} fields in the parsec-scale jets 
of the ``blue"
HBLs tend to be parallel to the local jet direction. This contrasts with the
tendency for the {\bf B} fields in the parsec-scale jets of the ``red" LBLs 
which are be 
perpendicular to the local jet direction.

\item The systematics of the magnetic field geometry thus do not
reflect the sequence in the SED peaks of CDQs, LBLs and HBLs.

\item The VLBI core polarization angles do not show any systematic
differences between HBLs and LBLs.

\item Several of the observed HBLs show evidence for a
``spine+sheath" magnetic-field structure, with transverse {\bf B} field 
in the inner region of the jet and longitudinal {\bf B} field at the edges.
This may point in the direction of a helical magnetic field threading the 
jets of these objects, in turn having implications for mechanisms
for the production of circular polarization in them.

\item Currently available circular polarization measurements in the literature
suggest that quasars may be more likely to show detectable circular
polarization than BL~Lacs. However, the situation is not clear, since the
largest difference in detection rates are found in the VLBI measurements, and
the low detection rate for the BL Lac objects in this case may be due to the
fact that the observed quasars have on average brighter VLBI cores than the
BL~Lacs.

\item Given the recent advances in circular polarization measurement
techniques on both arcsecond (ATCA) and milliarcsecond (VLBI) scales,
which have made detections possible in at least
a minority of  BL~Lacs (and quasars) with relatively faint compact
cores, it is beginning to become feasible to obtain circular polarization
measurements of well-defined samples on various scales. This would then
enable {\it rigorous} tests of the predictions of various scenarios for the
physical differences between different types of object. 

\end{enumerate}

\acknowledgements

We would like to acknowledge the contribution made by Ron Kollgaard and
Sally Laurent-Muehleisen in initiating the VLBP project on HEAO-1 BL Lacs,
without which this work would not have been possible. 
P. Shastri is grateful to the conference organisers for support that enabled 
her to make this presentation, and to the Department of Science \& Technology
and the Council for Scientific \& Industrial Research, Govt. of India, for
travel support.

\bibliographystyle{klunamed}
\bibliography{KharbShastri}

\begin{thebibliography}{}

\bibitem[\protect\citeauthoryear{{Blandford} and
  {K\"onigl}}{1979}]{BlandfordKonigl79}
{Blandford}, R.~D. and A. {K\"onigl}: 1979, `{Relativistic jets as compact
  radio sources}'.
\newblock {\em \apj} {\bf 232}, 34--48.

\bibitem[\protect\citeauthoryear{{Cawthorne} et~al.}{1993}]{Cawthorne93}
{Cawthorne}, T.~V., J.~F.~C. {Wardle}, D.~H. {Roberts}, and D.~C. {Gabuzda}:
  1993, `{Milliarcsecond Polarization Structure of 24 Objects from the
  Pearson-Readhead Sample of Bright Extragalactic Radio Sources. II.
  Discussion}'.
\newblock {\em \apj} {\bf 416}, 519--+.

\bibitem[\protect\citeauthoryear{{Cotton}}{1989}]{Cotton89}
{Cotton}, W.~D.: 1989, `{Very Long Baseline Interferometry, Techniques and
  Applications}'.
\newblock {\em eds. M. Fellini \& R.E. Spencer (Dordrecht :Kluwer)} p. 275.

\bibitem[\protect\citeauthoryear{{Fanaroff} and
  {Riley}}{1974}]{FanaroffRiley74}
{Fanaroff}, B.~L. and J.~M. {Riley}: 1974, `{The morphology of extragalactic
  radio sources of high and low luminosity}'.
\newblock {\em \mnras} {\bf 167}, 31P--36P.

\bibitem[\protect\citeauthoryear{{Gabuzda} et~al.}{2000}]{Gabuzda00}
{Gabuzda}, D.~C., A.~B. {Pushkarev}, and T.~V. {Cawthorne}: 2000, `{Analysis of
  {$\lambda$}=6cm VLBI polarization observations of a complete sample of
  northern BL Lacertae objects}'.
\newblock {\em \mnras} {\bf 319}, 1109--1124.

\bibitem[\protect\citeauthoryear{{Ghisellini} and
  {Maraschi}}{1989}]{Ghisellini89}
{Ghisellini}, G. and L. {Maraschi}: 1989, `{Bulk acceleration in relativistic
  jets and the spectral properties of blazars}'.
\newblock {\em \apj} {\bf 340}, 181--189.

\bibitem[\protect\citeauthoryear{{Homan} et~al.}{2001}]{HomanAttridgeWardle01}
{Homan}, D.~C., J.~M. {Attridge}, and J.~F.~C. {Wardle}: 2001, `{Parsec-Scale
  Circular Polarization Observations of 40 Blazars}'.
\newblock {\em \apj} {\bf 556}, 113--120.

\bibitem[\protect\citeauthoryear{{Homan} and {Wardle}}{1999}]{HomanWardle99}
{Homan}, D.~C. and J.~F.~C. {Wardle}: 1999, `{Detection and Measurement of
  Parsec-Scale Circular Polarization in Four AGNS}'.
\newblock {\em \aj} {\bf 118}, 1942--1962.

\bibitem[\protect\citeauthoryear{{Kollgaard} et~al.}{1996}]{Kollgaard96}
{Kollgaard}, R.~I., D.~C. {Gabuzda}, and E.~D. {Feigelson}: 1996,
  `{Parsec-Scale Radio Structure of Four X-Ray--selected BL Lacertae Objects}'.
\newblock {\em \apj} {\bf 460}, 174--+.

\bibitem[\protect\citeauthoryear{{Komesaroff} et~al.}{1984}]{Komesaroff84}
{Komesaroff}, M.~M., J.~A. {Roberts}, D.~K. {Milne}, P.~T. {Rayner}, and D.~J.
  {Cooke}: 1984, `{Circular and linear polarization variations of compact radio
  sources}'.
\newblock {\em \mnras} {\bf 208}, 409--425.

\bibitem[\protect\citeauthoryear{{Kuehr} and {Schmidt}}{1990}]{KuehrSchmidt90}
{Kuehr}, H. and G.~D. {Schmidt}: 1990, `{Complete samples of radio-selected BL
  Lac objects}'.
\newblock {\em \aj} {\bf 99}, 1--6.

\bibitem[\protect\citeauthoryear{{Laurent-Muehleisen}
  et~al.}{1993}]{LaurentMuehleisen93}
{Laurent-Muehleisen}, S.~A., R.~I. {Kollgaard}, G.~A. {Moellenbrock}, and E.~D.
  {Feigelson}: 1993, `{Radio morphology and parent population of X-ray selected
  BL Lacertae objects}'.
\newblock {\em \aj} {\bf 106}, 875--898.

\bibitem[\protect\citeauthoryear{{Laurent-Muehleisen}
  et~al.}{1997}]{LaurentMuehleisen97}
{Laurent-Muehleisen}, S.~A., R.~I. {Kollgaard}, P.~J. {Ryan}, E.~D.
  {Feigelson}, W. {Brinkmann}, and J. {Siebert}: 1997, `{Radio-loud active
  galaxies in the northern ROSAT All-Sky Survey. I. Radio identifications}'.
\newblock {\em \aaps} {\bf 122}, 235--247.

\bibitem[\protect\citeauthoryear{{Orr} and {Browne}}{1982}]{OrrBrowne82}
{Orr}, M.~J.~L. and I.~W.~A. {Browne}: 1982, `{Relativistic beaming and quasar
  statistics}'.
\newblock {\em \mnras} {\bf 200}, 1067--1080.

\bibitem[\protect\citeauthoryear{{Rayner} et~al.}{2000}]{Rayner00}
{Rayner}, D.~P., R.~P. {Norris}, and R.~J. {Sault}: 2000, `{Radio circular
  polarization of active galaxies}'.
\newblock {\em \mnras} {\bf 319}, 484--496.

\bibitem[\protect\citeauthoryear{{Sambruna}}{1994}]{Sambruna94}
{Sambruna}, R.: 1994, `{Ph.D. thesis, SISSA, Trieste}'.

\bibitem[\protect\citeauthoryear{{Sambruna} et~al.}{1996}]{Sambruna96}
{Sambruna}, R.~M., L. {Maraschi}, and C.~M. {Urry}: 1996, `{On the Spectral
  Energy Distributions of Blazars}'.
\newblock {\em \apj} {\bf 463}, 444--+.

\bibitem[\protect\citeauthoryear{{Stocke} et~al.}{1985}]{Stocke85}
{Stocke}, J.~T., J. {Liebert}, G. {Schmidt}, I.~M. {Gioia}, T. {Maccacaro},
  R.~E. {Schild}, D. {Maccagni}, and H.~C. {Arp}: 1985, `{Optical and radio
  properties of X-ray selected BL Lacertae objects}'.
\newblock {\em \apj} {\bf 298}, 619--629.

\bibitem[\protect\citeauthoryear{{Urry} and {Padovani}}{1995}]{UrryPadovani95}
{Urry}, C.~M. and P. {Padovani}: 1995, `{Unified Schemes for Radio-Loud Active
  Galactic Nuclei}'.
\newblock {\em \pasp} {\bf 107}, 803--+.

\bibitem[\protect\citeauthoryear{{Wardle} et~al.}{1984}]{Wardle84}
{Wardle}, J.~F.~C., R.~L. {Moore}, and J.~R.~P. {Angel}: 1984, `{The radio
  morphology of blazars and relationships to optical polarization and to normal
  radio galaxies}'.
\newblock {\em \apj} {\bf 279}, 93--98.

\bibitem[\protect\citeauthoryear{{Weiler} and {de
  Pater}}{1983}]{WeilerDePater83}
{Weiler}, K.~W. and I. {de Pater}: 1983, `{A catalog of high accuracy circular
  polarization measurements}'.
\newblock {\em \apjs} {\bf 52}, 293--327.

\bibitem[\protect\citeauthoryear{{Wood} et~al.}{1984}]{Wood84}
{Wood}, K.~S., J.~F. {Meekins}, D.~J. {Yentis}, H.~W. {Smathers}, D.~P.
  {McNutt}, R.~D. {Bleach}, H. {Friedman}, E.~T. {Byram}, T.~A. {Chubb}, and M.
  {Meidav}: 1984, `{The HEAO A-1 X-ray source catalog}'.
\newblock {\em \apjs} {\bf 56}, 507--649.

\end{thebibliography}
\end{article}
\end{document}